# Atomistic studies of thin film growth

Talat S. Rahman[1*], Chandana Ghosh[1], Oleg Trushin[2], Abdelkader Kara[1], Altaf Karim[1]
[1]Department of Physics, Cardwell Hall, Kansas State University, Manhattan, KS 66506
[2]Institute of Microelectronics and Informatics, Russian Academy of Sciences,
Yaroslavl 150007, Russia.
*email:rahman@phys.ksu.edu

## ABSTRACT

We present here a summary of some recent techniques used for atomistic studies of thin film growth and morphological evolution. Specific attention is given to a new kinetic Monte Carlo technique in which the usage of unique labeling schemes of the environment of the diffusing entity allows the development of a closed data base of 49 single atom diffusion processes for periphery motion. The activation energy barriers and diffusion paths are calculated using reliable manybody interatomic potentials. The application of the technique to the diffusion of 2-dimensional Cu clusters on Cu(111) shows interesting trends in the diffusion rate and in the frequencies of the microscopic mechanisms which are responsible for the motion of the clusters, as a function of cluster size and temperature. The results are compared with those obtained from yet another novel kinetic Monte Carlo technique in which an open data base of the energetics and diffusion paths of microscopic processes is continuously updated as needed. Comparisons are made with experimental data where available.

## 1. INTRODUCTION

One of the challenges in recent studies of materials at the nanoscale is the development of an understanding of microscopic processes that control thin film growth. This is a necessary task if we are to build materials of choice by design. It is also a daunting task because a faithful study demands seamless integration of information obtained at the microscopic level into formulations which predict and characterize behavior of systems at the macroscopic scale. We are speaking here of differences of many orders of magnitude. Phenomena at the atomic level extend themselves over nanometers with characteristic time scales of femto ($10^{-15}$) or pico ($10^{-12}$) seconds, while thin films for industrial applications are of mesoscopic (~microns) or macroscopic (>millimeter) dimensions and typically take milli-seconds or more to grow and evolve morphologically. Multiscale modeling which has become popular these days remains as yet a challenge, although thanks to advances in computational methodologies accompanied by the availability of intriguing experimental data, the field is advancing fast. To date theoretical studies of thin film growth proceed along one of several disparate approaches. A good deal of theoretical studies of thin films have been based on macroscopic approaches in which the films are treated as elastic solids suitable for application of formalisms of continuum mechanics [1,2]. Such models are capable of reproducing macroscopic properties of thin films starting from some basic assumptions about the processes responsible for them. On the other hand, models based on mean field theory and rate equations [3] make more explicit reference to microscopic processes through scaling laws and their comparison with experimental data. While quantities like diffusion coefficients and adsorption energies appear in the rate equation approach, these models are not yet capable of incorporating microscopic spatial information in their formulation. In recent times a number of hybrid models like configurational continuum [4] and level set method [5] have also been proposed. Provided appropriate microscopic parameters are available, these recent developments may provide a methodology for emulating the behavior of thin films over a large range of length and time scales. For details of some of the achievements of these hybrid models and their future prospects and challenges, the reader is referred to a recent review article by Ratsch and Venebles [6].

At the other end of the spectrum of multiscale modeling of thin films, fundamental studies are being carried out at the atomistic level using as accurate a technique as feasible. In combination with techniques like kinetic Monte Carlo (KMC) these microscopic models are also expected to facilitate simulation of thin film growth for realistic length and time scales. These microscopic studies are critical because of the experimentally demonstrated impact that structural and vibrational properties at the atomic level have on the eventual quality and properties of thin films. For example, whether a film grows layer-by-layer, or through the formation of 3D islands, depends on the details of the motion of adatoms on the potential energy surface provided by the substrate. Three types of growth modes are often discussed in the literature. The Frank-van der Merwe or layer-by-layer growth, and the Volmer-Weber or 3D island growth, appear to be accompanied by the more complex Stranski-Krastanov mode in which a competition between the other two types exists. The simple explanation of the first two types was provided by Schwoebel [7] and Ehrlich [8] who proposed that the existence (or lack thereof) of an additional activation energy barrier as an atom tries to decend a step edge, could be deciding factor for a 3D or layer-by-layer evolution of the film under growth conditions. This is the so-called "Schwoebel/Ehrlich" barrier whose determination from theory and experiments has led to substantial clarity in understanding thin film growth. The existence of the Stranski-Krastonov mode, however, implies that thin film growth patterns may be far more complex in general, and may require consideration of the role of quantities like surface strain and local perturbations.

Since the subject of atomistic studies of thin film growth is itself quite vast, our focus in this paper will be on some aspects of modeling of post-deposition evolution of thin film morphology through the diffusion of adatom and atomic clusters on surfaces. The objective will be to provide the reader a flavor of the types of issues encountered and available computational and theoretical methods to address them, rather than to present a full review of the subject. After highlighting in section II some of the basic ingredients needed for atomistic modeling of thin film growth, a summary of the theoretical methods is presented in section III. This is followed in section IV with some details of kinetic Monte Carlo simulations and its application to study cluster diffusion on metal surfaces. Some conclusions and thoughts for future directions are presented in section V.

## II. BASIC INGREDIENTS FOR ATOMISTIC MODELING

As the above paragraphs indicate, atomistic studies of thin film growth are replete with complex and competing events, each with its own characteristics. Some such processes in epitaxial growth depicted in Fig. 1, include adsorption (a), followed by the diffusion of the atom (called adatom) on the terrace (b), or its nucleation (c), or the attachment of an adatom to an existing island (d), or the reverse process of an adatom detachment from an existing island (e). In the same spirit, the adtom diffuses along a step edge (f), or down the step (g), or nucleate on top of an island (h). The diffusion of the dimer (i), as well as, that of clusters with larger number of atoms, may also proceed with significant rates. The nucleation of dimers, trimers, and other adatom and vacancy clusters themselves provide further avenues for anisotropic diffusion since the steps, edges, and corners formed by them may not be symmetric in geometry or in energetics. Stochastic processes like the fluctuations of step edges and dynamical processes which may dominate the relative stability of steps and other defects may offer other avenues for complex growth patterns. Realistic modeling of thin film growth has to account for these and other processes as they unveil themselves.

The basic ingredients in atomistic modeling of thin film growth are thus linked with those responsible for the characterization of the diffusion of adatoms, vacancies, and their clusters on surfaces with specific crystallographic orientations and marked with defects and other local environments. As we know, diffusion is a thermally activated process in which entities move on a temperature dependent, dynamical surface provided by the substrate. The diffusing entities vibrate about their equilibrium positions and occasionally overcome the energy barrier to move to another site of low occupation energy. To mimic thin film growth and the evolution of its morphology, we need first and foremost a tabulation of all possible diffusion pathways, and the probability (or rate) with which a particular path (or process) might be undertaken. One way to obtain such information is through molecular dynamics (MD) simulations. But, as we shall see, straight forward as the method is, it has drastic limitations which leave it uncompetitive for such studies, at the moment. Various types of accelerated schemes are currently being developed with the hope of forcing rare events to become less rare [9,10]. Several intuitive and heuristic methods are also being applied, as we shall see below.

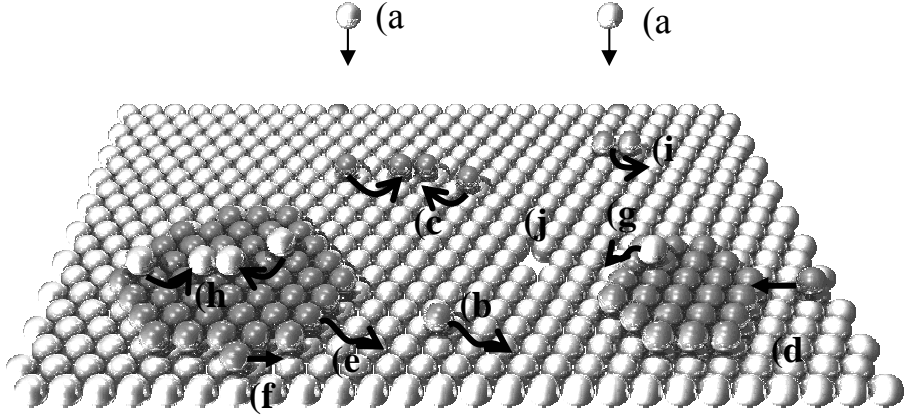

Fig. 1 Some atomic processes involved in epitaxial growth

For a given mechanism the diffusion rate is invariably obtained through the usage of transition state theory [which assumes that the process takes place through a well defined saddle point on the potential energy surface on which the diffusing entity is moving. The diffusion rate for process 'i' is then given by:

$$D_i = D_{0i} \exp(-\Delta E_i/k_B T). \tag{1}$$

where $\Delta E_i$ is the activation energy for the process, $k_B$ is Boltzmann constant, T is temperature, and $D_{0i}$ is the so-called pre-exponential or prefactor for the particular process. The above equation leads to the well known Arrhenius behavior. While we refer mostly to diffusion mechanisms in this paper, the above equation applies equally well to any process, with $D_i$ replacing the rate of the process. Thus foremost in the modeling of thin film growth is knowledge of mechanisms by which entities undergo spatial and temporal changes. In the case of diffusion this means knowledge of the actual paths for diffusion. Simple example of one such process is the hopping of an adatom from one equilibrium site to the next. Others may be more complex like diffusion via exchange of atoms or processes involving multiple atoms. Once diffusion mechanism and its path have been determined, the activation energy barrier is easily obtained from its definition as the difference of the total energy of the system at the maximum (saddle) and minimum (equilibrium) points along the path. The determination of the prefactors is less simple.

In Vineyard's work [11], the prefactor is given by the ratio of the products of the frequencies of the normal modes of the system with the diffusing entity in the minimum energy configuration, to those when it is at the saddle point in the diffusion path. Since in the latter configuration there is one less vibrational mode, $D_0$ has units of frequency and is often approximated by a value equivalent to that of a normal mode i.e. about $10^{12}$ or $10^{13}$ s$^{-1}$. Alternatively, the prefactor may be derived from the vibrational partition function, in which case it is given by the following expression:

$$D_{0i} = (k_B T/h)(nl^2/2\alpha) \exp(\Delta S_{vib}/k_B) \exp(-\Delta U_{vib}/k_B T). \tag{2}$$

where $\Delta S_{vib}$ and $\Delta U_{vib}$ are, respectively, the difference in the vibrational entropy and vibrational internal energy of the system with the diffusing entity at the saddle point and at the minimum energy configuration, h is Planck constant, l is the length of the jump, n the number of sites available for the jump, and $\alpha$ is the dimensionality. Equation 2 forms the basis for the recipe for calculations of prefactors for adatom diffusion on single crystal surfaces proposed by Kurpick et al. [12,13]. The thermodynamical quantities appearing in the above equations can be obtained through calculations of specific vibrational density of states of the system using standard lattice dynamical methods or MD simulations. Unusual values of prefactors may thus be expected if abnormal features appear in the vibrational density of states. However, as calculations of prefactors are tedious and time consuming, they have generally been given a constant numerical value, which is obviously a questionable assumption since several energetically competing process may differ substantially in their prefactors. Realistic values of prefactors and their impact on the dynamical evolution of thin films is a subject of recent investigations and beyond the scope of this work. We now turn to brief descriptions of some of the methods used to extract diffusions mechanisms, their paths and their activation energy barriers, and the methodology for examining dynamical evolution of thin films.

## III COMPUTATIONAL AND THEORETICAL METHODOLOGY

The first step in any atomistic calculation is a tractable and reliable procedure for describing the interatomic interactions and/or bonding. Energetics of the system including activation energy barriers, diffusion paths, and the potential energy surface can then be mapped out using standard techniques. The dynamical and spatial evolution of the system can next be examined through other sets of techniques. Some details of these methods are presented below.

### III.1 Methods for determining total energy

Theoretical techniques range from those based on first principles electronic structure calculations to those utilizing empirical or semiempirical model potentials. Although there are a variety of first principles or *ab initio* calculations familiar to physicists and chemists, for structural and dynamical studies of metal surfaces the current ones are generally based on density functional theory [14]. In this method, quantum mechanical equations are solved for the system electrons in the presence of ions/ion cores. The ion cores are allowed to relax to their minimum energy position, corresponding to 0 K, through calculations of the forces that act on them. Calculations of quantities like the total potential energy of the structure, equilibrium configurations of surface atoms (which are generally different from the bulk terminated ones), surface stress and surface energy are then performed with the ion cores at the minimum energy configuration corresponding to O K. These calculations are accurate and provide good insights into the electronic structural changes at surfaces that manifest themselves in a variety of forms.

First principles electronic structure calculations have received further boost in the recent past [15] with the introduction of new schemes [16] for obtaining solutions to the Kohn-Sham equations for the total energy of the system. One form of this technique solves also the equations of motion for the ion cores. This is the so-called first principles molecular dynamics method. Ideally the forces responsible for the motion of the ion-cores in a solid are evaluated at each step from self-consistent solutions to the Hamiltonian for the valence electrons. In practice, this makes the calculations very tedious. Since there are no fitting parameters in the theory, this approach has predictive power and is desirable for exploring the structure and the dynamics at any solid surface. There are, however, several obstacles both technical and conceptual in nature, which keep this method from broad applicability, for the moment. Until such *ab initio* methods become more feasible for realistic length and time scales, MD simulations of thin film diffusion processes will have to rely on model interaction potentials.

For these very reasons, a genre of many body interatomic potentials has found broad applications in the past few decades. These potentials have already provided good deal of information on the microscopic properties of a selected group of metal surfaces which been tested by comparison with experimental data. In what follows here we have also employed semi-empirical potentials based on the embedded atom method (EAM) [17]. These are many-body potentials and hence do not suffer from the unrealistic constraints that pair-potentials impose on the elastic constants and the vacancy formation energy. For the six fcc metals Ni, Pd, Pt, Cu Ag and Au, and their intermetallics, these potentials seem to have done an excellent job of reproducing many of the observed characteristics of the bulk metal and also of the surface systems  Another nice feature of these types of potentials is the ease with which they can be applied to rather large systems. Finally, once the potential has been fitted to the bulk properties no changes are made to obtain the surface properties. Hence at least for the surface phenomenon there are no free parameters. It should also be mentioned that there are several other realistic many-body potentials available. Our choice of EAM is based mostly on familiarity and easy access than a philosophical difference with the others.

### III.2 Methods for determining activation energy barriers and diffusion paths

Several new approaches have recently been proposed to overcome the limitations of length and time scales arising from standard MD simulations. Of these we provide some details here of one that we have used in our work. We have combined a simple activation technique, namely the Spherical Repulsion Minimization Method [18], with a tested method for locating the saddle point along the diffusion path as given by the Nudged Elastic Band (NEB) method [19] or the drag method. The entire procedure includes two stages. First, a repulsive potential is applied in the energy minimization stage (performed with atoms interacting via EAM potentials) to generate possible transition paths by which the system can escape from a particular energy minimum. Nontrivial transitions which avoid a return to the original point are thus obtained. Next, with well defined initial and final configurations, a reliable method is used for accurate determination of the activation energy barriers. Since the spherical repulsion method is relatively new we give some details below.

### III.2.1 Spherical Repulsion Minimization Method

The idea behind this method is to modify locally the energy surface of the system by applying a spherically repulsive potential which makes the specific minimum energy state unstable, but leaves the other nearby energy minima unaffected. Minimization of the total energy for the modified potential energy surface then activates the possible transitions for the system. To accomplish this, the initial configuration of interest is prepared by minimizing the total energy of the system using a standard technique like MD cooling. (In the MD cooling method, the energy is gradually minimized by setting the velocities (v) and the force F on a particle to satisfy the condition v.F < 0.) Next, the system is slightly displaced from the initial state by moving an active atom (one that is expected to diffuse) in the direction of the nearest available vacant site. Next the system is allowed to move to the other nearest minimum energy states by adding a localized repulsive potential to the Hamiltonian of the system of the form:

$$U_{tot}(r) = U(r) + Ae^{-(r-r_0)^2} \qquad (3)$$

where $r_0$ represents the coordinates of the initial state [18]. By varying the initial displacement and the form of the repulsive interaction, a range of final states of the diffusing entity may be generated. If the repulsive potential is sufficiently localized around the initial potential minimum, the final state energy can be made to depend only on the true potential of the system and not on the fictitious repulsive potential. The set of configurations needed to reach the final state from the initial state serves as the input to the calculation of the activation barriers from a method like NEB which we describe briefly below.

### III.2.2 Nudged Elastic Band Method

While the above repulsive potential minimization can be used to generate the final state configuration, it does not yield the minimum energy path (MEP) and the lowest activation barrier value for getting to this final state. For this purpose, we use the nudged elastic band method [19]. This is an efficient method for finding the MEP between a given initial and final state of a transition, given the knowledge of both initial and final states. The MEP is found by constructing a set of images of the system, in principle arbitrary, between the initial and final states. A spring interaction potential between adjacent images is added to ensure continuity of the path, thus mimicking an elastic band. The total force acting on an image is the sum of the spring force along the local tangent and the true force perpendicular to the local tangent [19]:

$$\vec{F}(i) = \vec{F}_i^s + \vec{\nabla} V(\vec{R}_i) \qquad (4)$$

where V is the total energy of the system and $R_i$ the set of atomic coordinates. The spring force is given by

$$\vec{F}_i^s = k_{i+1}(\vec{R}_{i+1} - \vec{R}_i) - k_i(\vec{R}_i - \vec{R}_{i-1}) \qquad (5)$$

where k is the spring constant. A minimization procedure for the force acting on the images should then bring images to the MEP. At any point along the path, the force acting on the particles points only along the path while the energy is constant for any degree of freedom in the direction perpendicular to the path. An initial guess for the images in the NEB is usually obtained by interpolating the particle configurations between the final and initial state. For the present application however, we find that this often leads to numerical instabilities due to the strong hard core repulsion of EAM potential. To circumvent this problem, we use the set of configurations generated in moving to the final state in the presence of the repulsive potential as the initial path. This leads to fast convergence in the NEB method without the instabilities encountered in the linear interpolation scheme.

**III.2.3 The drag method for determining energy barriers**

Although the NEB provides reliable and accurate procedure for the determination of diffusion barriers, it can be very slow. For our newly developed "self-teaching KMC" method [20] in which the energy barriers for the systems processes of choice are calculated simultaneously, we have found it prudent to invoke a faster method. One of the simplest of such is the drag method in which the diffusing atom is moved to an adjacent vacant site in small steps. At each step the diffusing atom is allowed to relax in the plane perpendicular to the direction of motion, while all other atoms are free to move in any direction. Since the neighboring adatoms are free to follow the diffusing one, this simple procedure is capable of activating many-particle processes. In our investigations of cluster diffusion, which we discuss below, we have found the drag method to provide activation energy barrier which are in good agreement with those obtained from the more sophisticated NEB method. The drag method may be further improved by including a fine grid over which the calculations of the energy barriers are done. Such a grid may also reveal diffusion paths that may not be accessible in a simple drag approach. An example below illustrates this point very clearly.

**III.2.4 Diffusion path and activation energy barrier from drag/grid method**

The combination of a fine grid with the traditional drag method is another powerful tool to unravel complex multi-particle mechanisms and their energetics. To illustrate the point, we consider the case of a descending adatom from a four-atom island on a fcc(111) surface, with all atoms in question being Ag as described by EAM potentials. In this particular scenario the Ag adtom prefers to diffuse by exchanging its position with a neighboring Ag atom, rather than by hoping to another site [21]. The adatom may exchange with one of the four atoms of the island forming, at the end of the process, a step with either a (100) or a (111)-microfacet (Fig.2a). The determination of the diffusion paths and energy activation barriers was done by calculating the total energy of the constrained system with the exchanged atom placed at all points on the grid and all other atoms allowed to relax. The exchanged atom is constrained in the surface plane but allowed to relax in the direction perpendicular to the surface. The collection of the total energy of the system evaluated at the grid points is then used to create a contour plot from which one can extract the minimum energy paths, saddle points and energy minima. In Fig. 2b we show the grid, the minimum energy paths, and the corresponding energy barriers. Note that in the lower grid, the system first encounters a metastable configuration with the exchanged atom occupying an hcp site. Note also that for both mechanisms, the diffusion path is not a straight line and a simple method like the drag may miss the minimum energy path.

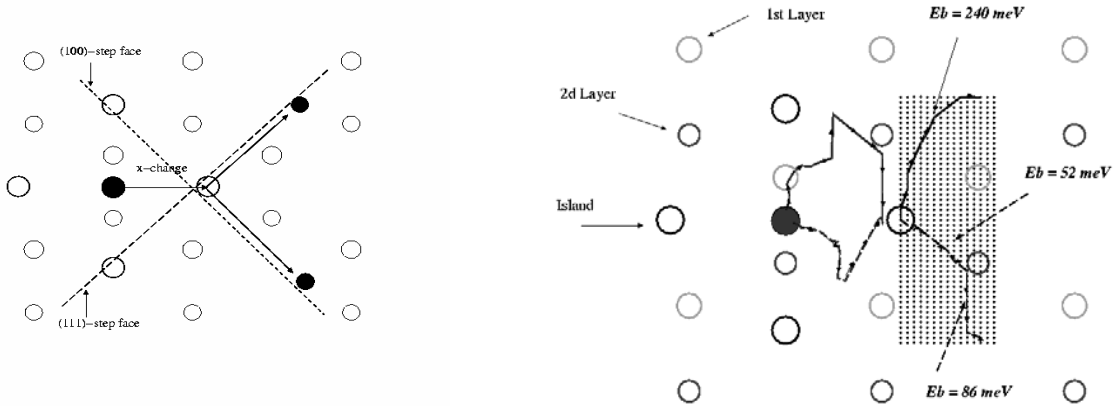

Fig. 2. Adatom decent from a small island: a) two possible scenarios; b) their paths and their energetics.

### III.3 Methods for simulating evolution of thin film morphology

There is a hierarchy of models for simulating epitaxial growth on spatial and temporal scales ranging from atomic to macroscopic. At the atomic level molecular dynamics or molecular static methods have been applied with interactions based on *ab-initio* or semiempirical approaches. On the other hand, information obtained at the atomic level (energy barriers, and pre-factors) is used in Kinetic Monte-Carlo simulations for simulating thin film growth and morphological evolution. We present below some details of these methods.

### III.3.1 Molecular dynamics technique

In molecular dynamics (MD) technique classical equations of motion, for atoms interacting with a known interatomic potential, are solved numerically using a suitable algorithm. The MD cell generally consists of a few thousand atoms arranged in ten to twenty layers. The minimum size of the cell depends on the nature of the dynamical property that one is interested in investigating. For atomistic simulations of thin films, periodic boundary conditions are applied in the two directions parallel to the surface while no such constraint is imposed in the direction normal to the surface. An algorithm like Nordsieck's with a time-step of $10^{-15}$s is then used to solve Newton's equations for all the atoms in the MD cell. For any desired temperature a preliminary simulation is carried out under conditions of constant temperature and constant pressure (NVT) to obtain the lattice constant at that temperature. The system with a particular surface crystallographic orientation is then generated in the bulk terminated positions. Under conditions of constant volume and constant temperature (NVT) this surface system is next equilibrated to the desired temperature. Next the system is allowed to evolve in a much longer run of a few ns, with its energy maintained as constant (microcanonical ensemble). Statistics on the positions and velocities of the atoms are recorded. Structural and dynamical properties of the system can now be obtained from appropriate correlation functions involving atomic positions and velocities. If forces are calculated from *ab initio* methods, MD simulations are capable of providing details of all atomistic processes in thin film growth. However, as we have already noted, it would take a lot of computer time to span length scales from $10^{-15}$s to $10^{-3}$s or more. It will also generate more data than one can absorb and unless creative methods are developed for extracting important information from the pile, it is not clear how viable this technique will be for developing an understanding of material properties from microscopic considerations.

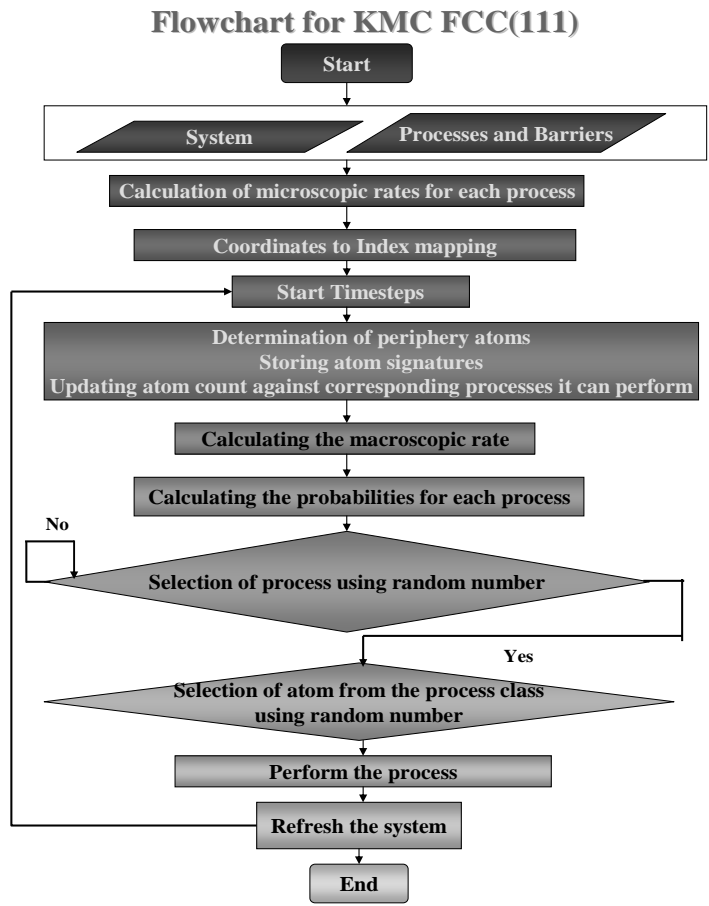

Figure 3. Flowchart for a kinetic Monte Carlo simulation

### III.3.2 The Kinetic Monte Carlo method

An alternative to MD simulations in examining surface phenomena is offered by the kinetic Monte-Carlo technique in which the rates of various eligible atomic processes are provided as input. If this input is reasonable and complete then KMC simulations may be carried out for time scales as in experiments. One of the challenges thus is to provide a complete set of atomistic processes which may be probable. Since this task is non-trivial, standard KMC simulations are performed with only a few processes as input and all others either ignored or included in approximate ways (e-g bong counting models). In recent times several efforts have been made to overcome this deficiency, in particular in the work of Henkelman and Jonsson [10].

The goal of kinetic Monte Carlo (KMC) is to mimic real experiments through sophisticated simulations. For simulations to be realistic, increasingly complex scenarios need to be implemented. At the heart of a KMC simulation of the time evolution of thin film morphology lie mechanisms that are responsible for determining the microscopic events to be performed at any given time. The rates at which these events can be performed are ultimately related to the time it takes for the system to evolve. If for diffusion process i, there are $n_i$ adatoms/entities capable of undergoing the process, the macroscopic rate associated with event (i) is simply $R_i = n_i D_i$, where $D_i$ is an defined in Eq. 1 . The total rate R for evolution of the system is then a sum over all possible events:

$$R = \sum_i R_i$$

In KMC simulations, at each step the acceptance of the chosen event is set to one. The probability that a given event is chosen is, however, dictated by its relative rate $P_i = R_i/R$, which a diffusing entity is randomly chosen from the set $\{n_i\}$ to perform. Classification of the active atoms into classes of the events that they can perform can be very time consuming. We have adopted, for our simulations of phenomena relevant to epitaxial growth, two novel pattern recognition schemes some details of which we give later. In Fig. 3, we show a typical flowchart of a KMC simulation as applied to fcc(111) crystals.

## IV. APPLICATION OF KMC TO CLUSTER DIFFUSION

Experimental studies of the diffusion of adatoms and small atomic clusters on metal surfaces using Field Ion Microscopy (FIM) have already provided a number of unexpected events such as the concerted motion of atoms [22] and the collective sliding motion of clusters [23]. These studies, together with the availability of reliable computational techniques, motivated a number of computational and theoretical scientists to examine in more detail the possible paths for the diffusion of adatom and vacancy clusters on metal surfaces and calculate their energetics and dynamics. Rapid developments in the technique of Scanning Tunneling Microscopy (STM) have further exposed interesting characteristics of the dynamics of adatom and vacancy islands of nm size. STM measurements on Ag(100) [24] have established that large adatoms move and those on Ag(111) [25] have confirmed that the mobility of the vacancy islands is comparable to that of the adatom islands. These and related observations led to a series of papers [26-29] with speculations about the microscopic mechanisms that cause these large islands to move. Of particular interest here are the competing mechanisms of adatom periphery diffusion, evaporation/condensation, and terrace diffusion. Statistical mechanical calculations based on solid-on-solid (SOS) models predict specific scaling of the diffusion coefficient with the island diameter, depending on the preponderance of one of these three mechanisms. Since these dependencies are not unequivocally extracted from experimental data, because of the large error bars involved, the issue is not yet completely settled, although the bias is towards periphery diffusion. Molecular dynamics simulations of Ag vacancy island on Ag(111) [30], on the other hand, have shown a preference for periphery diffusion , while in the case of the elastic-continuum models the three mechanisms lead to characteristic decays of the time and space correlation functions. Several questions about the elastic-continuum based models, however, remain as shown by Bogicevic *et al* [31] who find that the exponents in the power law dependence of the diffusion coefficient on island size were themselves temperature dependent and material specific, unlike predictions of the simpler earlier model calculations. While the work of Bogicevic *et al* points to the simplicity of the previous calculations, it also begs the question whether the atomistic model based on a few hand-picked atomistic processes is capable of displaying the inherent complexity of the system. The issue is whether the evolution of the system could be prejudiced by the usage of an insufficient set of atomic processes arising from a narrow local consideration.

The main point of departure in our work is the usage of pattern recognition schemes in KMC simulations which allows the creation of data bases containing most processes that the system under investigation might require. We have achieved this goal through two schemes as relevant to the diffusion of adatoms on fcc(111) . In the first we have chosen a "minimum configuration" scheme in which we record, from the 36 atoms in the 3 shells surrounding an active or "central" atom, only those sites (vacant or occupied) that will *uniquely* determine the process that is associated with a particular atom. In Fig. 4, as an example, we show the minimum configuration associated with diffusion along step A. Step A signifies a (100)-microfacetted step while step B signifies a (111)-microfacetted step. For this process, the conditions to be met for adatom diffusion are: i) sites 4, 5, 17 *must* be filled (Black) and ii) sites 0, 1, 2, 3, 6, 7 *must* be vacant (White). The rest of the sites could be either filled or vacant (grey).

This schemes is particularly easy to operate and in the application to the diffusion of adatom clusters on fcc(111) surface, forty-nine basic processes (and their equivalents obtained by applying the symmetry operations for a hexagonal lattice) are found for single atom peripheral diffusion [32]. A summary of the activation energy barriers for all processes demanded by long KMC runs for 2D island diffusion through single atom periphery motion are summarized in Table III. The calculations were performed using interaction potentials from EAM and diffusion paths from NEB. In the other scheme, the configurations with all atoms in the three shells surrounding an active atom were given unique labels and the spherical repulsion method described above together with NEB and drag methods were used to generate all possible final configurations and their energetics. These were then saved in a data base which was kept open and updated as subsequent simulations generated new processes whose activation energy barriers were calculated on the fly [20]. Simulations were performed with the open data base until the system evolution reached equilibrium conditions, as judged by a count of nearest neighbor bonds of the active atoms.

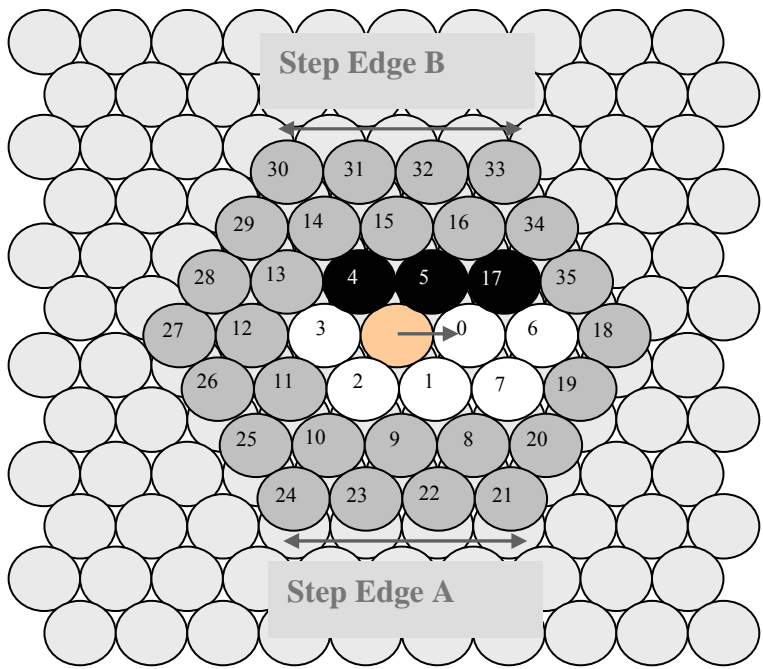

Fig. 4. Three shells of nearest neighbor atoms around the central atom (atom selected to move) and the pattern recognition scheme.

**IV.1 Some results from cluster diffusion**

We present here some results of simulations of the diffusion of 2D Cu islands on Cu(111), containing 10-1000 atoms, for about 500 million MC-steps at several temperatures. In these we have used the closed data-base approach with 49 processes forming a complete set of single-atom peripheral motion. Since the physical time elapsed at each MC-step is governed by the rate of the process, they are unequal in length. Thus, to calculate the mean square displacement of the center of mass, we filter our data to a set of almost equidistant MC-step along with the corresponding center of mass coordinates. From this new filtered data set we calculate the mean square displacement using correlations between these time intervals. These plots clearly show a linear behavior, as can be seen in Fig. 5 and Fig. 6 for clusters containing 19 and 100 atoms, respectively.

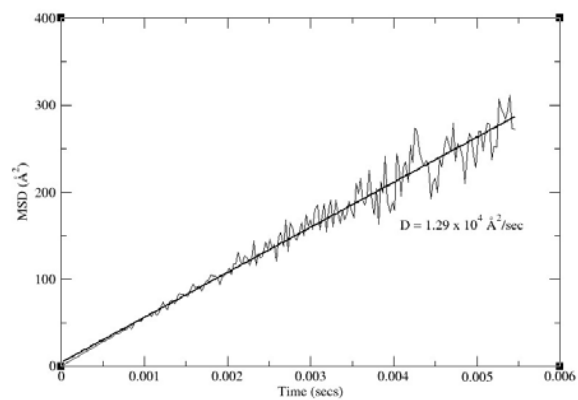

Fig. 5. Mean square displacement of the CM for a 19 atoms at 500K

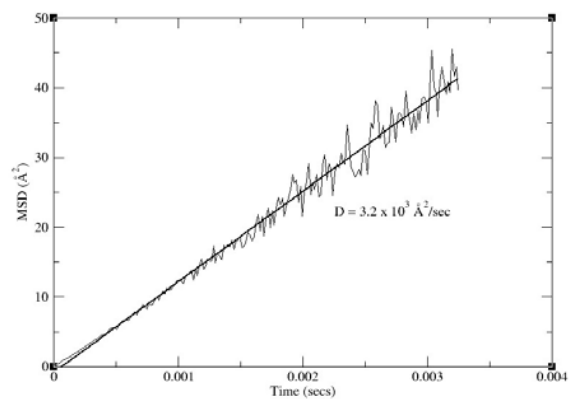

Fig. 6 Mean square displacement of the CM for a 100 atoms at 500K

The diffusion coefficients of clusters of various sizes calculated at three temperatures is summarized in Table I. As compared to a 10 atom cluster, we find that the diffusion coefficient of a 100 atom cluster to be three orders of magnitude slower at temperatures of 500K and 1000K.

Table I. Diffusion coefficient for different clusters at different temperatures

| Cluster size (atoms) | Diffusion Coefficient D ($Å^2$/sec) | | |
| --- | --- | --- | --- |
| | Temperature of simulation | | |
| | 300 K | 500 K | 1000 K |
| 10 | 40 | $1.12 \times 10^6$ | $1.62 \times 10^9$ |
| 14 | 6.0 | $2.0 \times 10^5$ | $4.35 \times 10^8$ |
| 18 | 3.8 | $1.09 \times 10^5$ | $2.32 \times 10^8$ |
| **19** | - | **$1.29 \times 10^4$** | **$3.08 \times 10^8$** |
| **21** | - | **$9.49 \times 10^3$** | **$1.59 \times 10^8$** |
| 26 | 0.48 | $2.24 \times 10^4$ | $1.54 \times 10^8$ |
| **38** | - | **$2.22 \times 10^4$** | **$7.61 \times 10^7$** |
| **100** | - | **$3.2 \times 10^3$** | **$4.7 \times 10^6$** |

The frequencies of all the processes performed during 100 million MC steps are reported in Table II, together with the energy barriers calculated for the process. The Table shows clearly the dependence of the frequencies of events on the simulation temperature. There is a dependence on the cluster size and shape, but it is not as dramatic as that on the temperature. The example of statistics for the 19-atoms island in Table II, displays the asymmetry in the diffusion along the two types of steps on Cu(111), particularly at lower temperatures.

**Table II** Frequency of Processes for the 19 atom cluster (Hexagon) at three temperatures

| Temperature / Processes | Energy Barrier (eV) | 300K | 500K | 1000K |
|---|---|---|---|---|
| Step Edge A | 0.252 | **0.62** | **0.42** | **0.23** |
| Step Edge B | 0.295 | **0.17** | **0.24** | **0.15** |
| Kink Detach along Step A | 0.519 | 0.0 | 0.0020 | **0.019** |
| Kink Detach along Step B | 0.556 | 0.0 | 0.0 | **0.012** |
| Kink Detach along Step (small) A | 0.608 | **0.026** | **0.012** | **0.043** |
| Kink Detach along Step (small) B | 0.680 | 0.0016 | 0.0023 | **0.019** |
| Kink Incorp. A | 0.220 | 0.0 | 0.0020 | **0.018** |
| Kink Incorp. B | 0.265 | 0.0 | 0.0 | **0.011** |
| Kink Incorp. (small) A | 0.0075 | **0.025** | **0.011** | **0.037** |
| Kink Incorp. (small) B | 0.0810 | 0.0 | 0.0012 | **0.012** |
| Kink Detach out of Step A | 0.658 | 0.0 | 0.0 | 0.0013 |
| Kink Detach out of Step B | 0.590 | 0.0 | 0.0 | 0.0027 |
| Kink Fall into Step A | 0.074 | 0.0 | 0.0 | 0.0016 |
| Kink Fall into Step B | 0.0069 | 0.0 | 0.0 | 0.0023 |
| Kink Rounding A | 0.656 | 0.0 | 0.0 | 0.0011 |
| Kink Rounding B | 0.678 | 0.0 | 0.0 | 0.0 |
| KESE A | 0.374 | 0.0 | 0.0011 | 0.0078 |
| KESE B | 0.402 | 0.0 | 0.0 | 0.0066 |
| Corner Rounding at AA stage 1 | 0.313 | 0.0 | 0.0 | **0.014** |
| Corner Rounding at AA stage 2 | 0.143 | 0.0 | 0.0 | 0.0050 |
| Corner Rounding at AA stage 3 | 0.0096 | 0.0 | 0.0 | **0.012** |
| Corner Rounding at BB stage 1 | 0.374 | 0.0 | 0.0 | 0.0079 |
| Corner Rounding at BB stage 2 | 0.038 | 0.0 | 0.0 | **0.012** |
| Corner Rounding at BB stage 3 | 0.052 | 0.0 | 0.0 | 0.0049 |
| Corner Rounding at AB stage 1 | 0.317 | **0.066** | **0.11** | **0.11** |
| Corner Rounding at AB stage 2 | 0.0839 | 0.0053 | **0.024** | **0.055** |
| Corner Rounding at BA stage 1 | 0.396 | 0.0047 | **0.023** | **0.051** |
| Corner Rounding at BA stage 2 | 0.0148 | **0.067** | **0.12** | **0.11** |
| Rounding a chain A | 0.0626 | 0.0 | 0.0 | 0.0053 |
| Rounding a chain B | 0.0191 | 0.0 | 0.0 | 0.0054 |

The Arrhenius plot drawn for the diffusion of the clusters, in Fig. 7, is interesting since it yields values of their effective energy barriers which may also be extracted from experiments. From our simulations the 10, 14, 18 and 26 atom clusters have effective barriers of 0.64 eV, 0.66 eV, 0.65 eV and 0.71 eV, respectively. There appears to be a slight increase in the effective barrier for the 26 atom cluster as compared to that for the 10, 14 and 18 atom clusters which have almost the same value. We need to carry out further investigations of the microscopic details of the diffusion processes to draw definite conclusion in this regard.

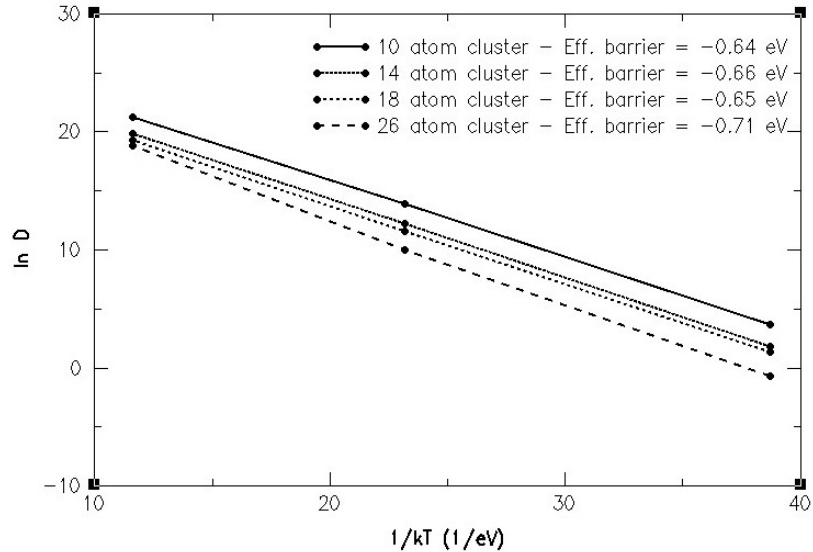

Figure 7. Arrhenius plot with the effective barriers for the 10 atom, 14 atom, 18 atom and 26 atom clusters

Some insights into the microscopic details of the cluster size and temperature dependencies of the diffusion processes may be obtained from the frequencies of the various events as recorded in our KMC simulations. A common feature for clusters is that diffusion along step edge A dominates that along step edge B. Of course, this also implies larger occurrence of step A as compared to step B, the rationale for which is in turn related to the corner rounding processes. Several subtle differences can be found for the frequencies of processes involving kink incorporation, corner rounding, etc., for the four types of clusters considered. Interestingly simulations carried out with the open data base, present several scenarios which are not in agreement with those obtained with the closed data base. Although the number of multiple atom processes are not found to be large (less than 1%), there are subtle differences, particularly arising from the attachment and detachment of atoms from the cluster that account for some of the differences. As a result the scaling of the diffusion coefficient with cluster size is found to be significantly different in the two cases. We are investigating the reasons for this difference.

**IV.2 Some results from island coalescence**

As an example of application of the KMC simulation with the open data-base (Self Teaching KMC), we present here results of the coalescence process in which two adatom islands join together to form a larger island with an equilibrium shape on Cu(111). Successive snapshots of the system during KMC simulation are shown in Fig 8. This is a remarkable result as our simulations show almost perfect agreement with the experimental observations of Giesen et al.[33]. Note that in these simulations the system was free to evolve with the diffusion mechanisms of its choice.

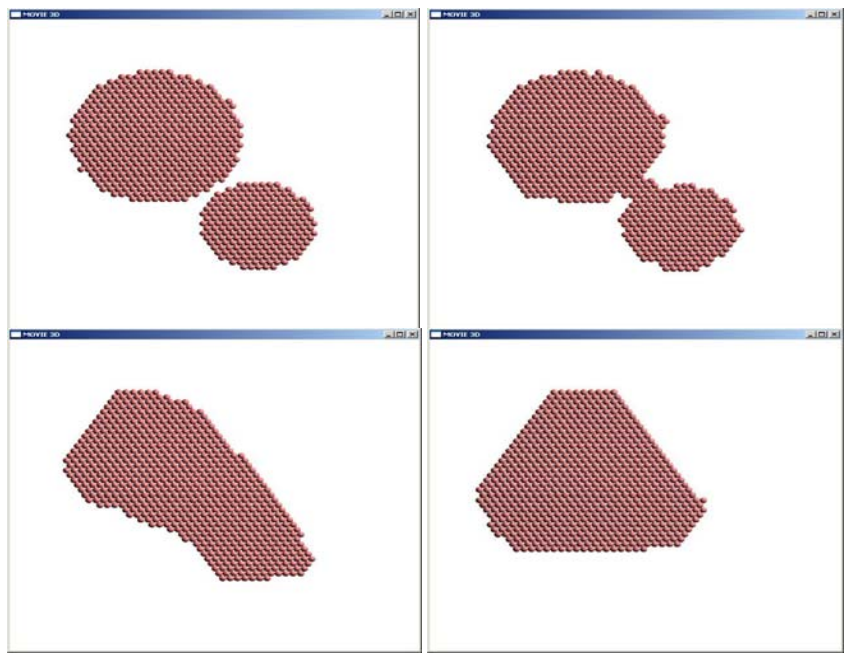

Fig. 8 Results of KMC simulation of the coalescence of two islands performed at 400K

## V. CONCLUSIONS

The above sections provide a brief summary of some of the techniques that are used in atomistic modeling of thin film growth and its morphological evolution. The field is still in its infancy as accurate methods like *ab initio* electronic structure calculations are only now becoming feasible for systems with as much complexity as those presented here. Once activation energy barriers of all relevant processes and their diffusion paths can be obtained from such methods, KMC simulations appear to provide an attractive procedure for predicting and understanding the characteristics of thin films as a function of their atomistic structure, substrate crystallography, and temperature. As we have already alluded to, the task of calculating diffusion prefactors is still ahead of us. This is particularly important since we find many competing processes to differ only slightly in energy and differences in their vibrational entropy contributions to the prefactors can make a difference in the ultimate evolution of the film morphology. Another important result from our simulations with the open data base is that dynamical evolution of the system with prejudged diffusion processes may yield erroneous results. Also, the pattern recognition schemes to be a prudent way to develop data base of diffusion processes and their energetics. It does involve a lot of work in the beginning but once the data base is compiled, it can be used for any type of simulation of the system. Of course, for realistic simulations of thin films we need to incorporate exchange and other processes which involve motion in 3D. Such effort is currently underway.

## ACKNOWLEDGEMENTS

This work was supported partially by a grant from NSF (ERC-0085604) and from CRDF.

## REFERENCES

1. J. Villain, J. Phys. **11**, 19 (1991).
2. J. Krug, Adv. Phys. **46**, 139 (1997)
3. J. A. Venables, Philos. Mag. 27 697 (1973)
4. N. Israeli and D. Kandel, Phys. Rev. Lett. **88**, 116103 (2002)
5. C. Ratsch, A. Zangwill, P. Smilauer, and D. D. Vvedensky, Phys. Rev. Lett. **72**, 3194 (1994)
6. C. Ratsch and J. A. Venables, J. Vac. Sci. Technol. A 21, S96 (2003)
7. R. L. Schwoebel, and E. J. Shipsey, J. App. Phys. **37**, 3682 (1966).
8. G. Ehrlich, and F. G. Hudda, J. Chem. Phys. **44**, 1039 (1966).
9. A. F. Voter, Phys. Rev. Lett. **78**, 3908 (1997)
10. G. Henkelman and H. Johnson Phys. Rev. Lett. **90**, 116101 (2003)
11. G. H. Vineyard, J. Phys. Chem. Solids **3**, 121 (1957)
12. U. Kürpick, A. Kara and T.S. Rahman, Phys. Rev. Lett. **78,** 1086 (1997).
13. U. Kürpick and T.S. Rahman, Phys. Rev. B **57**, 2482 (1998).
14. P. Hohenberg and W. Kohn, Phys. Rev. B **136**, 864 (1964).
15. For a review see: M.C. Payne, M.P. Teter, D.C. Allen, T.A.Arias and J.D. Joannopoulos, Rev. Mod. Phys. **64**, 1045 (1992).
16. R. Car and M. Parrinello, Phys. Rev. Let.55, 2471 (1985).
17. S. M. Foiles, M. I. Baskes, and M. S. Daw, Phys Rev. B **33**, 7983 (1986).
18. O. Trushin, E. Granato, S.C. Ying, P. Salo, T. Ala-Nissila, Phys. Rev. B **65**,.241408 (2002).
19. H. Jonsson, G. Mills and K. W. Jacobsen, in *Classical and Quantum Dynamics in Condensed Phase Simulations*, ed. by B. J. Berne *et al* World Scientific, Singapore, 1998.
20. O. Trushin, A. Kara, and T. S. Rahman, to be published
21. A. Karim, A. Kara, A. Al-Rawi and T. S. Rahman, "Diffusion Paths, Barriers and Prefactors: Ag clusters on Ag(111)," in *Collective Diffusion on Surfaces:Correlation Effects and Adatom Interactions*, edts M.C.Tringides and Z. Chvoj (Kluwer 2001).
22. G. L. Kellog and A. F. Voter, Phys. Rev. Lett., **67** 622, (1991).
23. S. C. Wang, U. Kürpick, and G. Ehrlich, Phys. Rev. Lett. **81**, 4923 (1998).
24. W.W. Pai, A.K. Swan, Z. Zhang, and J.F. Wendelken, Phys. Rev. Lett. **79**, 3210 (1997).
25. K. Morgenstern, G. Rosenfeld, B. Poelsema, and G. Comsa, Phys. Rev. Lett. **74**, 2058, (1995);K. Morgenstern, G. Rosenfeld, and G. Comsa, Phys. Rev. Lett. **76**, 2113, (1996).
26. J.M. Soler, Phys. Rev. B **50**, 5578 (1994).
27. J.M. Wen, S.L. Chang, J.W. Burnett, J.W. Evans, and P.A. Thiel, Phys. Rev. Lett. **73**, 2591 (1994).
28. J.M. Soler, Phys. Rev. B **53**, R10 540 (1996).
29. S. V. Khare, N. C. Bartelt, and T. L. Einstein, Phys. Rev. Lett. **75**, 2148 (1995); J. C. Hamilton, M. S. Daw, and S. M. Foiles, Phys. Rev. Lett. **74**, 2760 (1995); C. DeW. Van Siclen, Phys. Rev. Lett. **75**, 1574 (1995).
30. U. Kurpick, P. Kurpick, and T. S. Rahman, Surf. Sci. Lett. **383**, L713 (1997).
31. A. Bogicevic, S. Liu, J. Jacobsen, B. Lundqvist, and H. Metiu, Phys. Rev. B **57**, R9459-R9462 (1998).
32. C. Ghosh, A. Kara, and T. S. Rahman, to be published; C. Ghosh. PhD thesis, Kansas State University, 2003.
33. M. Giesen, private communication